# Improving data quality for 3D electron diffraction (3D ED) by Gatan Image Filter and a new crystal tracking method

**Taimin Yang[a*], Hongyi Xu[a*], Xiaodong Zou[a*]**

a Department of Materials and Environmental Chemistry (MMK), Stockholm University, Svante Arrhenius väg 16C, Stockholm, SE-10691, Sweden

* Correspondence email: taimin.yang@mmk.su.se, hongyi.xu@mmk.su.se, xiaodong.zou@mmk.su.se

**Abstract**

3D ED is an effective technique to determine the structures of submicron- or nano-sized crystals. In this paper, we implemented energy-filtered 3D ED using a Gatan Energy Filter (GIF) in both selected area electron diffraction mode and micro/nanoprobe mode. We explained the setup in detail, which improves the accessibility of energy-filtered 3D ED experiments as more electron microscopes are equipped with a GIF than an in-column filter. We also proposed a crystal tracking method in STEM mode using live HAADF image stream. This method enables us to collect energy-filtered 3D ED datasets in STEM mode with a larger tilt range without foregoing any frames. In order to compare the differences between energy-filtered 3D ED and normal 3D ED data, three crystalline samples have been studied in detail. We observed that the final $R_1$ will improve 20% to 30% for energy-filtered datasets compared with unfiltered datasets and the structure became more reasonable. We also discussed the possible reasons that lead to the improvement.

## 1. Introduction

Over the last three decades, 3D electron diffraction (3D ED) has been developed into a reliable technique for structure determination, which is complementary to single-crystal X-ray diffraction (SCXRD), powder XRD and cryo-EM single particle analysis. The development of computerized 3D ED as a method for structure determination is pioneered by electron diffraction tomography (EDT) (Kolb *et al.*, 2007) and rotation electron diffraction (RED) (Wan *et al.*, 2013), which utilize stepwise rotation (goniometer + electron beam used for RED) along a single axis. More recently, new protocols, such as continuous rotation MicroED (Nannenga *et al.*, 2014; Rodriguez *et al.*, 2015; Jones *et al.*, 2018; Lanza *et al.*, 2019), fast-EDT(Gemmi *et al.*, 2015; Plana-Ruiz *et al.*, 2020) and continuous rotation electron diffraction (cRED) (Seo Seungwan *et al.*, 2018; Wang, Yang *et al.*, 2018; Cichocka *et al.*, 2018; Xu *et al.*, 2018; Wang *et al.*, 2019; Xu *et al.*, 2019) have been developed for determining crystal structures of beam-sensitive materials. They are based on continuously rotating the sample stage at constant speed while collecting ED patterns simultaneously. To stabilize the crystals in the vacuum environment, a cryo-holder or a cryo-microscopy is usually used. These techniques are widely applied in structure determination of proteins (Nederlof *et al.*, 2013; Clabbers *et al.*, 2017; Clabbers & Abrahams, 2018), zeolites (Gemmi *et al.*, 2015; Lee *et al.*, 2018), metal-organic frameworks (MOFs)(Cichocka *et al.*, 2020), covalent-organic frameworks (COFs) (Liu *et al.*, 2016) and small organic molecules (Gruene *et al.*, 2018).

The strong interaction between electrons and materials also brings problems, for example, dynamical scattering. With kinematic approximation used in most of the current 3D ED protocols, lots of structural details can be revealed already, such as ions, water and ligand bindings to a protein molecule (Clabbers *et al.*, 2020; Clabbers & Xu, 2020), hydrogen atoms in small organic compounds(Clabbers *et al.*, 2019), partial atomic occupancies, modulated structures (Li *et al.*, 2020) and organic structure templating agents (OSDAs) in MOF (Wang, Rhauderwiek *et al.*, 2018). However, usually there are still some unexplainable difference electrostatic potential maps. These residue potentials may hinder further discovery of finer structures during refinement and the final $R_1$ value is usually larger than 10%, which will consider questionable for many small molecule X-ray crystallographers. Palatinus et.al. applied EDT with precession (PEDT) and dynamical refinement to model dynamical scattering (Palatinus, Petříček *et al.*, 2015; Palatinus, Corrêa *et al.*, 2015). With their technique, they were able to find the positions of all the hydrogen atoms in paracetamol crystals(Palatinus *et al.*, 2017). In addition, dynamical refinement can be utilized for the determination of absolute crystal structures (Brázda *et al.*, 2019). Currently, dynamical refinement is a very calculation intensive process, even for small unit cells. Furthermore, this technique relies upon PEDT, which requires longer data acquisition time.

Another issue for electron diffraction is inelastic scattering, which increases the background level in diffraction patterns. This background is most obvious in electron diffraction patterns at low angles. Even though modern diffraction data software (XDS, DIALS, MOSFLM) has sophisticated background removal algorithms to deal with this, the existence of inelastic scattering will still add errors in the diffraction experiment. The inelastically scattered electrons can be removed by an energy filter. Yonekura et.al. utilized an in-column omega energy filter and did a systematic investigation on charges in protein crystals (Yonekura *et al.*, 2015; Yonekura & Maki-Yonekura, 2016; Yonekura *et al.*, 2018, 2019; Maki-Yonekura *et al.*, 2020). Gemmi et.al reported energy filtered 3D ED and they found the obtained structure from filtered datasets was closer to the results of X-ray refinement (Gemmi & Oleynikov, 2013). However, these experiments require an in-column omega energy filter, which is not commonly equipped in transmission electron microscopes (TEMs).

Here, we implemented energy-filtered 3D ED using a Gatan Energy Filter (GIF) in both selected area electron diffraction (SAED) and micro/nanoprobe mode. Nowadays, most TEMs are equipped with a GIF, making the method developed in our work widely available for researchers. The main advantage for this method is removing the inelastically scattered electrons, which might remove a part of dynamical scattered electrons at the same time(Yonekura *et al.*, 2015). In addition, in order to track the crystal in scanning electron transmission electron microscopy (STEM) mode, we developed a tracking method based on monitoring live HAADF image stream. This can avoid crystal moving out of the beam during the tilting and the tilt range can always reach the maximum tilt range of the microscope (in our case ~150°).

## 2. Experimental Section

### 2.1. Samples

Datasets from NaCl, $NH_4H_2PO_4$, and finned zeolite ZSM-5 crystals (Dai *et al.*, 2020) were collected in during the experiments. In total 8, 6 and 2 datasets were acquired for NaCl, $NH_4H_2PO_4$ and ZSM-5 crystals, respectively and half of them were energy filtered datasets. The samples were crushed in a mortar and dispersed in ethanol. A drop of suspension was dipped to a lacey-carbon copper grid.

### 2.2. Data collection

All the datasets were collected on a Themis Z double aberration-corrected microscope. The microscope is equipped with a monochromator, which was used for adjusting the beam current draw from the electron gun. A Fischione model 2020 tomography holder was used. A GIF Quantum with an UltraScan1000FX CCD camera was used for data collection. The width of the energy slit was adjusted to 10eV and zero-loss peak (ZLP) was selected in all experiments. Before each session, an automatic alignment provided by Digital Micrograph and manual alignment are required. The position of the slit in both X and Y direction needs to be checked manually. It is recommended that once the alignment is completed, the current of all lenses in the projection system should remain to be unchanged. For example, if the camera length is changed, the GIF has to be aligned again. Besides, the position of the energy slit should be checked occasionally to avoid slit position shift during the data collection session.

#### 2.2.1 Data collection in TEM mode

The schematic of energy-filtered 3D ED in TEM mode was shown in Figure 1 (a). The microscope was operated in TEM mode. In order to avoid influencing the alignment of GIF, we used the C3 aperture to define the illumination area on the sample instead of selected area aperture during the diffraction experiment. The size of C3 aperture is 30μm, which produces beam size of 800 nm. The benefit of doing this is during data collection, other crystals near the target crystal will not be illuminated. Since the alignment of GIF can be easily affected by the settings of intermediate lenses and projection lenses, it was impossible to use defocused diffraction patterns to track the crystal during data collection (Cichocka *et al.*, 2018). As a result, the tilting range was limited to -40° to +40°. In addition, the eucentric height must be calibrated carefully to minimize the sample shift during tilting.

#### 2.2.2 Data collection in STEM mode with crystal tracking

The schematic of energy-filtered 3D ED in STEM mode was shown in Figure 1 (b). In order to track the crystal during continuous rotation data collection, STEM HAADF imaging can be used. In this protocol, the microscope was operating in STEM mode and the beam was scanned in an area smaller than the crystal size, in order to obtain live HAADF image stream. The beam scan and HAADF image collection were controlled and visualized by Velox software from Thermo Fisher Scientific. The electron beam can be set in the micro-probe mode or in the nano-probe mode. The size of C2 aperture was 50μm, which is the smallest C2 aperture available in our microscope. In the micro-probe mode, the convergence angle of the beam was adjusted to 0.5 mrads and then the current of C3 lens was tuned to make the beam parallel in order to obtain sharp reflections. The diameter of the parallel electron beam was around 350nm. In the nano-probe beam mode, the minimum convergence angle was 5 mrads. Parallel beam and sharp spots can be obtained by adjusting the current of C3 lens. The diameter of the probe was estimated to 300nm. During the data collection, the beam can be kept stationary, however the tilt range will be limited because the crystal can easily move out of the beam at high tilt angle. Therefore, we propose the scan the beam over a small area in order to form a STEM image of the crystal for tracking. During beam scan, the diffraction pattern will have some small movements due to slight beam tilt when the scan area is large, which will blur the whole diffraction pattern. By keeping the scan area relatively small (~200nm), the movement can be reduced for maintaining sharp reflections. Meanwhile, a HAADF detector was used to collect electrons scattered to high angles and track the position of the crystal. The detailed workflow for STEM HAADF crystal tracking is illustrated in Figure 2. After alignment of the microscope, a HAADF image was collected at 5000x magnification with a parallel beam of 300 nm in size for identifying the position of possible crystals, as shown in Figure S1. Then the target crystal was moved to the center and live HAADF image stream was started. At the same time, the data collection was started and the stage began to rotate. During the rotation, if the target crystal moved out of the beam, one edge of the blurry HAADF image will change in contrast (i.e. become darker), as shown in Figure S2(a). To put the crystal back into the beam, we need to move the crystal towards the dark area using the joystick, as shown in Figure S2(b). Meanwhile, the rest of electrons not participate in the STEM HAADF imaging will go through the GIF and forming energy-filtered diffraction patterns on the camera. As the crystal is being continuously rotated, energy filtered 3D ED data was collected with live crystal tracking by HAADF imaging.

The 3D ED data collection was done by using the data acquisition software *Instamatic*. The rotation speed of the goniometer was kept at 0.3°/s. The exposure time was 1s per frame for all datasets. The dose rate of the incident beam for all experiments was adjusted to 0.1 e/Å$^2$/s.

### 2.3. Data processing and structure determination

After data collection, the datasets were processed by XDS for spot finding, unit-cell determination, indexing, space-group assignment, data integration, scaling, refinement and generating Shelx HKL files (Kabsch, 2010). Data statistics indicators provided in the output CORRECT.LP file were used for data quality comparisons. Next, ShelxT (Sheldrick, 2015*b*) was used for structure solution. Structure refinement and visualization of the structure models by were by ShelxL (Sheldrick, 2015*a*) and ShelXle (Hübschle *et al.*, 2011).

## 3. Results and Discussion

### 3.1. Energy-filtered 3DED in STEM mode

We first performed 3D ED experiments in STEM mode, with a parallel beam of size around 300nm. With a nano-sized probe, we can keep the illuminated area as small as possible, avoid damaging the nearby crystals. This setup can significantly increase the dose efficiency compared with the SAED setup (typical diameter of illuminated area is around 2-4μm). By collecting energy-filtered 3D ED in STEM mode, there is no need to switch between imaging mode and diffraction mode, which is another advantage over operating in TEM mode. Because of inherent hysteresis of the projection lenses, the alignment of the GIF will drift away after switching back and forth between imaging and diffraction mode. Therefore, most of the experiments reported in this work were done in STEM mode.

#### 3.1.1. NaCl

Because of the simple structure and exceptional crystallinity of NaCl, it was chosen as a testing sample to explore the experimental and refinement parameters, showing the improvement in data and structural quality by using energy filtered 3D ED. Table S1 shows the XDS data processing results for NaCl crystals. Due to the adaption of the live STEM-HAADF tracking method, the tilt range for all the datasets reached above 130°. Two datasets even reached 150°, which is the maximum tilt range for the holder and microscope combination. The tilt range for other datasets were smaller because of blockage by neighboring crystals or the copper grid. There were some distortions in the obtained unit cell parameters in this acquisition configuration and the distortion was due to the GIF. The data processing was performed in space group $P1$. Otherwise, positions of the predicted spots will deviate from the observed spots. Improvement in final $R_1$ values were observed using energy filtered 3D ED data, from 13.9% to 8.4% in average, as shown in Table S1.

### 3.1.2 Refinement with EXTI keywords

During the refinement of the NaCl structure, we introduced the keyword “EXTI” in the SHELX input file and we found it had a huge impact on the final $R_1$ values for all 8 datasets. In the structure of NaCl, the position of the Na atom and Cl atom will not change during refinement. Only the atomic displacement parameters (ADPs) and scale factors will change. Without the EXTI keyword, the final $R_1$ value will be around 20-35%. After using EXTI to weight the reflections, the final $R_1$ value will decrease sharply to around 9%. Although the number of refined parameters increased from 3 to 4, this huge decrease cannot be explained simply by an additional parameter. Furthermore, unlike structures refined with the EXTI keyword, in which the ADPs of all atoms were normal, negative ADPs were found in the structures refined without the EXTI keyword, as shown in Figure S3(b). To prove the better $R_1$ is not the result of overfitting, we tried the “SWAT” keyword and the number of refined parameters increased to 5. However, the final $R_1$ values were only slightly improved (around 20-30%) with some ADPs being negative, as shown in Figure S3(c).

As shown in the refinement results from all three samples, EXTI can significantly improves the final $R_1$ value and the structure model, while SWAT keyword cannot, even though this keyword adds one more refined parameter than EXTI. EXTI is mainly used for correcting extinction effect of X-ray (Chandrasekhar, 1960), which does not exist in electron diffraction. According to the SHELX manual (https://shelx.uni-goettingen.de/shelxl_html.php), when EXTI is added in the input file, the calculated intensity is modified as:

$$F_c^2(new) = F_c^2 * \frac{k^2}{\sqrt{1 + \frac{0.01 F_c^2 \lambda^3 x}{\sin(2\theta)}}} \quad (1)$$

$k$ is the overall scale factor and $x$ is the refined parameter of EXTI

SWAT is the keyword to model diffuse solvent. While SWAT is added in the input file, the calculated intensity becomes:

$$F_c^2(new) = F_c^2 * \left(1 - g * \exp\left[-8\pi^2 U \left(\frac{sin\theta}{\lambda}\right)^2\right]\right)^2 \quad (2)$$

Where g and U are refined parameters of SWAT keyword.

The most obvious difference between these two corrections is the EXTI correction will make the relationship between the observed intensity and the calculated structure factor amplitude closer to linear while after SWAT correction the relationship between the observed intensity and calculated structure factor amplitude will still be quadratic. The linear relationship between the observed and calculated intensity is first proposed by Darwin (M.A, 1914*b*,*a*). He calculated the theoretical diffracted intensity of X-ray for perfect crystal using dynamical theory and found the linear relationship between observed intensity and the structure factor amplitude (Chandrasekhar, 1960):

$$\rho = \frac{8}{3\pi}\frac{Ne^2\lambda^2}{mc^2}|F|\frac{1 + |\cos 2\theta|}{2 \sin 2\theta} \quad (3)$$

In this formula, N is the number of unit cells per unit volume. F is the structure factor corrected for thermal vibration. $\theta$ is the Bragg angle.

On the other hand, the quadratic relationship is only valid when the whole crystal is composed of many slightly misoriented crystal domains (mosaic). The reflections from different blocks are optically independent and therefore the total intensity is the sum of the intensities from individual blocks. Each block is in itself perfect but its volume $\Delta v$ is so small that its integrated intensity for one block is given by:

$$\rho = N^2\lambda^3\left(\frac{e^2}{mc^2}\right)^2 |F|^2 \frac{1 + cos^2 2\theta}{2 \sin 2\theta} * \Delta v \quad (4)$$

Because all three samples were highly crystallized and do not contain any solvents in the structure, they are more likely to act like a perfect crystal in the electron beam.

The EXTI formula also indicates the level of calculated structure factor correction is different across different resolution shell and structure factors. When the intensity is higher, larger $F_c^2$ will also make the term $(0.01F_c^2\lambda^3x)/\sin(2\theta)$ more significant compared with 1, thus making the relationship closer to linearity. At lower resolution, smaller $\sin(2\theta)$ will also make the relationship closer to linearity. This is also consistent with the observations of D. Dorset (Dorset, 1995). Reflections with high intensity and at low resolution make more contribution to the process of dynamical scattering. As a result, for all samples, we applied EXTI keywords to partially compensate for the dynamical scattering.

### 3.1.3. $NH_4H_2PO_4$

Next, we collected six $NH_4H_2PO_4$ datasets (3 unfiltered and 3 energy-filtered). Similar improvement of data quality and structure refinement statistics have been observed. For these crystals, the tilt range was around 140° and the averaged final $R_1$ value decreased from 12.4% (1.5%) to 10.2% (1.3%) using energy-filtered 3D ED data, as shown in Table 1. Similar distortion of lattice was observed for all datasets and typical 3D reciprocal lattice was visualized by REDp software. As shown in Figure 3, the angles between the axes were deviated from 90 degree (in the tetragonal crystal). The deviation was likely caused by the elliptical distortion from projection lenses or GIF lenses. We also compared the structure refined from six datasets. For all three energy-filtered datasets, all hydrogen atoms in the structures could be located and the bond angle between hydrogen atoms and other atoms were reasonable. After anisotropic refinement, all the ADPs were positive and the shape of the ellipsoid are chemically sensible, as shown in Figure 4 (a). Anisotropic displacement parameters are known to act as indicators for poor quality data if the displacement ellipsoids conflict with prior chemical knowledge. As a proof of data quality improvement, we showed the anisotropic refined structures in Figure 4 (b) and (c) as a comparison. The refinement for unfiltered datasets was not able to find all the hydrogen atoms around the nitrogen atom or unable to obtain reasonable thermal parameters for the nitrogen atom.

### 3.1.4. Evaluation of the stability of the new tracking procedure

We use $NH_4H_2PO_4$ datasets as examples to evaluate the stability of the new crystal tracking protocol. The normalized scaling factors from the tilt series determined by XDS (SCALE in file INIT.Lp) can be used to judge whether the crystal has moved out of the beam during data collection. The SCALE factor uses the ED frames only and is employed in XDS to correct for variations in the incident beam flux and diffraction volume of crystal. When the crystal moves out of the beam scanning area, the corresponding diffracted intensities will be lower and higher scale factor needs to be applied to that frame. Figure 5 shows the scale factor of filtered datasets and without filtered datasets, both applied the tracking method. The plot of the scale factor revealed a very smooth and slow variation profile over the whole tilt range, indicating the diffraction volume of the crystal is relatively stable and the method is effective for 3D ED data collection.

### 3.1.3. ZSM-5

The structures of both NaCl and $NH_4H_2PO_4$ crystals are quite simple and they are relatively stable compared with porous materials. Therefore, we used finned ZSM-5 as an example to show energy filtration will improve the data quality and structure refinement of more complex structure. A comparison between the refined structure of finned ZSM-5 was shown in Figure 6. In Figure 6 (a), the structure refined against unfiltered 3D ED data contains negative ADPs (shown in thin rectangles) and very thin ellipsoids. The oscillation direction of some ellipsoids was along the bond, which is chemically unreasonable. In contrast, the structure from filtered datasets was much better. The ADPs were very reasonable and no negative ADPs can be observed, as shown in Figure 6 (b). The data statistics and refinement results were summarized in Table 2. The I/SIGMA and CC(1/2) were improved in energy-filtered data and the final $R_1$ value also decreased from 0.264 to 0.243. In addition, the average deviation from the reference structure was compared in Table S4. Both averaged deviations of Si atoms and oxygens were reduced by around 0.01Å using the energy-filtered data. We also compared the bond length of both filtered and unfiltered structure with the reference structure and we found the bond length for unfiltered structure has some exceptional short Si-O bond, as highlighted in red in Table S5, while the bond length of filtered structure narrower spread.

## 3.2. Energy-filtered 3D ED of ZSM-5 in TEM mode

3D Data was collected on the same ZSM-5 crystal over the same tilt range with and without energy filtration. The data processing results were summarized in Table S6. The most significant improvement in data quality is the increase of I/SIGMA, from 7.74 for unfiltered dataset to 9.25 for energy filtered dataset. In addition, CC(1/2) was improved a little, from 0.997 to 0.998. However, the tilt range was limited at around 70° because of sample movement during tilting. It is quite difficult to use the defocused image tracking method as implemented in *Instamatic* (Cichocka *et al.*, 2018) because the alignment of GIF needs to be adjusted after changing the current of the intermediate lens and it is quite hard to realign the lenses of GIF within 1s. Therefore, it is necessary to put the crystal close to the mechanical eucentric height.

### 3.3. Comparison with other methods

There are two possible factors that contributed to the improvement of data quality. The first one is the removal of background noise (as shown in Figure S4), so that the data processing software can extract the intensity more accurately. However, even without the energy filter, the low angle reflections were not submerged in the 'tail' of the direct beam, unlike the case for protein crystals. Therefore, another possible reason is inelastically scattered electrons contain a part of dynamically scattered electrons. When the electrons interact with materials, some electrons lose energy after the scattering events while the rest were elastic scattered. The more times an electron interacts with the crystal, there is a higher probability for the electrons to scatter inelastically, which will be removed by the energy filter, thus alleviating the influence of dynamical scattering (Yonekura *et al.*, 2015). However, we suspect that only a limited portion of the dynamically scattered electrons was removed.

To our knowledge, this is the first time that energy-filtered 3D ED data were collected using a GIF. Previously, Yonekura *et. al.* and Gemmi *et. al.* performed energy-filtered 3D ED experiments using an in-column omega filter(Gemmi & Oleynikov, 2013; Yonekura *et al.*, 2015). The GIF system is much more widely available comparing to the in-column filter, making the method developed in this work applicable in more TEM labs.

Kolb *et.al* developed automated diffraction tomography (ADT), which can also track crystals during tilting (Kolb *et al.*, 2019). However, their technique requires a pre-recorded STEM image tilt series from a fiducial marker or the target crystal itself, the drift of the target during tilting can be calculated and compensated by shifting the electron beam. This technique will also require switching between focused STEM beam and quasi-parallel STEM beam. Compare with ADT, our live STEM-HAADF tracking protocol can allow data collection over larger tilt range. Furthermore, the method developed in this work also reduces the overall electron dose and data acquisition time as no pre-recording of the TEM/STEM images were needed.

Compared with the crystal tracking method using defocused diffraction pattern(Cichocka *et al.*, 2018), the STEM-HADDF live tracking does not sacrifice any frames to form shadow images. Therefore, the completeness of the dataset is higher for data collected over the same tilt range. Another advantage is our method checks the position of the crystal very frequently. Usually, we can adjust the scanning speed, the number of pixels and the dwelling time and control the time for acquiring each STEM image at 1 to 2 s. In contrast, the defocused diffraction pattern method will show the position of the crystal every 10 or 20 frames and each frame takes around 0.5-2s, as shown in Figure S5 (a) and (b). The operator needs more experience for performing crystal tracking with such a long interval since the crystal may move out of the beam already before the next defocused image was formed.

There are some challenges for energy-filtered 3D ED experiments. The first challenge is the distortion of crystal lattice brought by the energy filter. Sometimes the distortion can be so large that it is difficult to use the correct space group during XDS data processing. Another challenge for energy-filtered 3D ED is finding the target crystals. As shown in Figure S1, because of parallel illumination, weak beam and large collection angle, if the crystals are small, the HAADF image can be so blurry that it is hard to spot these crystals.

## 4. Conclusion

In this work, we described the implementation of energy-filtered 3D ED setup using a GIF, in both TEM and STEM mode in detail. Furthermore, we proposed a live crystal tracking method using STEM-HAADF image stream to keep the target crystal in the electron beam over large tilt range. Using this crystal tracking method, it is possible to collect 3D ED data close to 150°, which is the maximum tilt range in our holder-TEM setup. Compared with crystal tracking by occasionally defocusing the diffraction beam, no electron diffraction frames are lost in the 3D ED dataset. In addition, we collected multiple datasets from NaCl and $NH_4H_2PO_4$ crystals and found an improvement in final $R_1$ value after energy filtration. The obtained structure after refinement also improved. In the $NH_4H_2PO_4$ case, all hydrogen atoms were found and the atomic displacement parameters were chemically sensible when energy-filtered data were used. The improvement of data quality opens up new possibilities for studying atomic motion, disorder and charge distribution from submicrometre-sized crystals.

At this stage, the energy-filtered 3DED data collection still requires an operator to correct the stage position according to the contrast of HAADF image. In the future, we hope to use this feature to achieve automated data collection with live crystal tracking. With the increased interests in radiation-sensitive materials, high level of automation is a way to reduce electron dose on a sample and to improve throughput of structure determination. The crystal tracking method proposed here are generally applicable and can be applied to all types of GIF/camera setups.

Movies of the data collection using crystal tracking and the crystallographic data for both structures in CIF format are provided as supporting information. The datasets collected in this study have been deposited at zendo.

Deposition Number 2046826 contains the supplementary crystallographic data for this paper. These data are provided free of charge by the joint Cambridge Crystallographic Data Centre and Fachinformationszentrum Karlsruhe Access Structures service www.ccdc.cam.ac.uk/structures.

**Authorship contribution statement**

**Taimin Yang**: Conceptualization, Investigation, Implementation, Programming, Data acquisition, Data analysis, Manuscript writing and revision. **Hongyi Xu**: Supervision, Manuscript revision. **Xiaodong Zou**: Supervision, Manuscript revision.

**Conflicts of interest**

The authors declare that they have no known competing financial interests or personal relationships that could have appeared to influence the work reported in this paper.

**Acknowledgement**

The authors thank Dr. Thomas Thersleff EFTEM training and nice suggestions. The authors thank Mr. Linus Schönström for his inspiring DM script and discussions. The authors thank Mrs.

Yangjun Liu for helpful and inspiring discussions. The authors thank Prof. Jeffrey Rimer and Dr. Heng Dai for providing finned ZSM-5 sample.

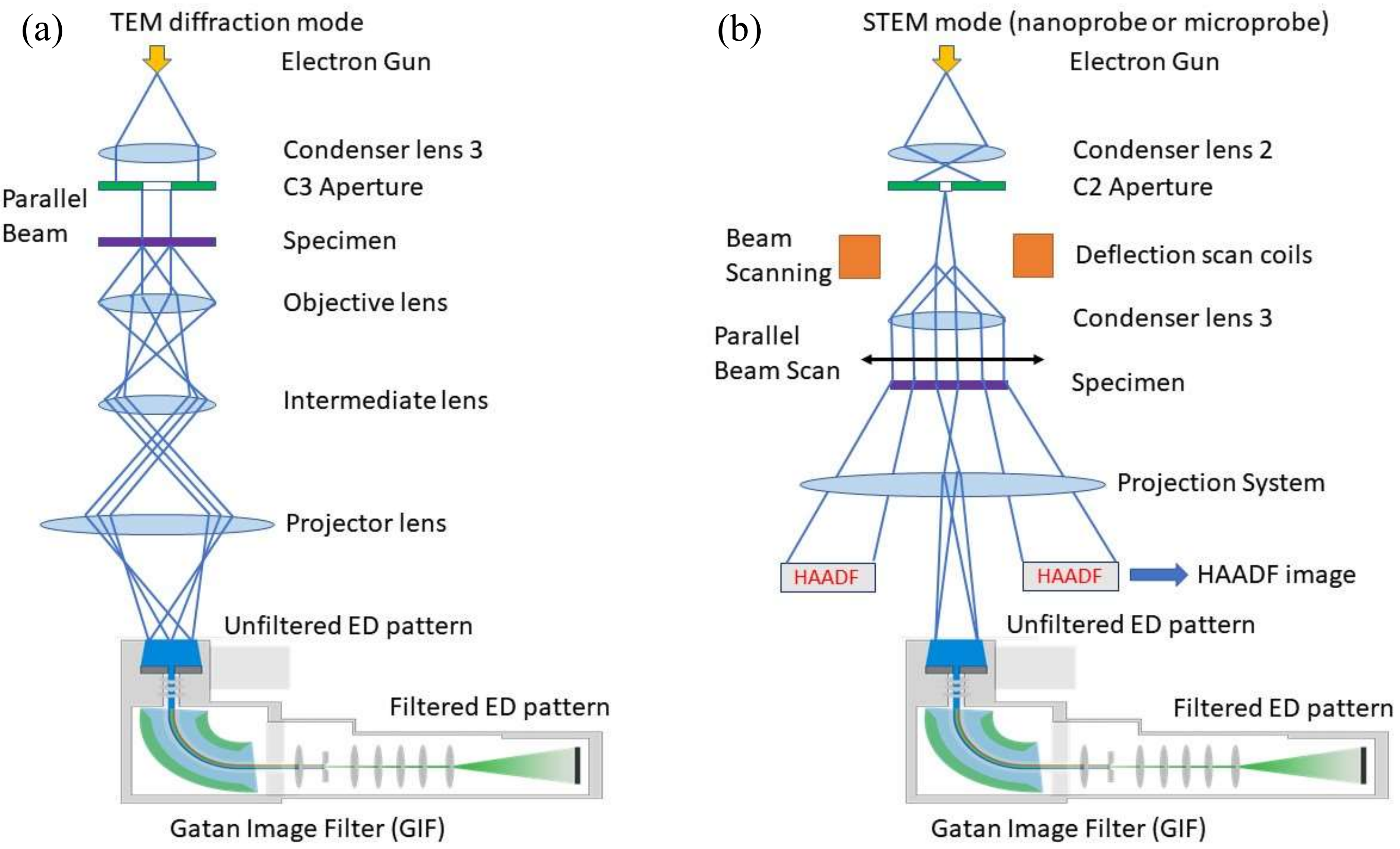

Figure 1. Implementation of Energy filtered 3D ED in (a) TEM mode (b) STEM mode. In STEM mode, crystal tracking using live STEM-HAADF image stream can be activated.

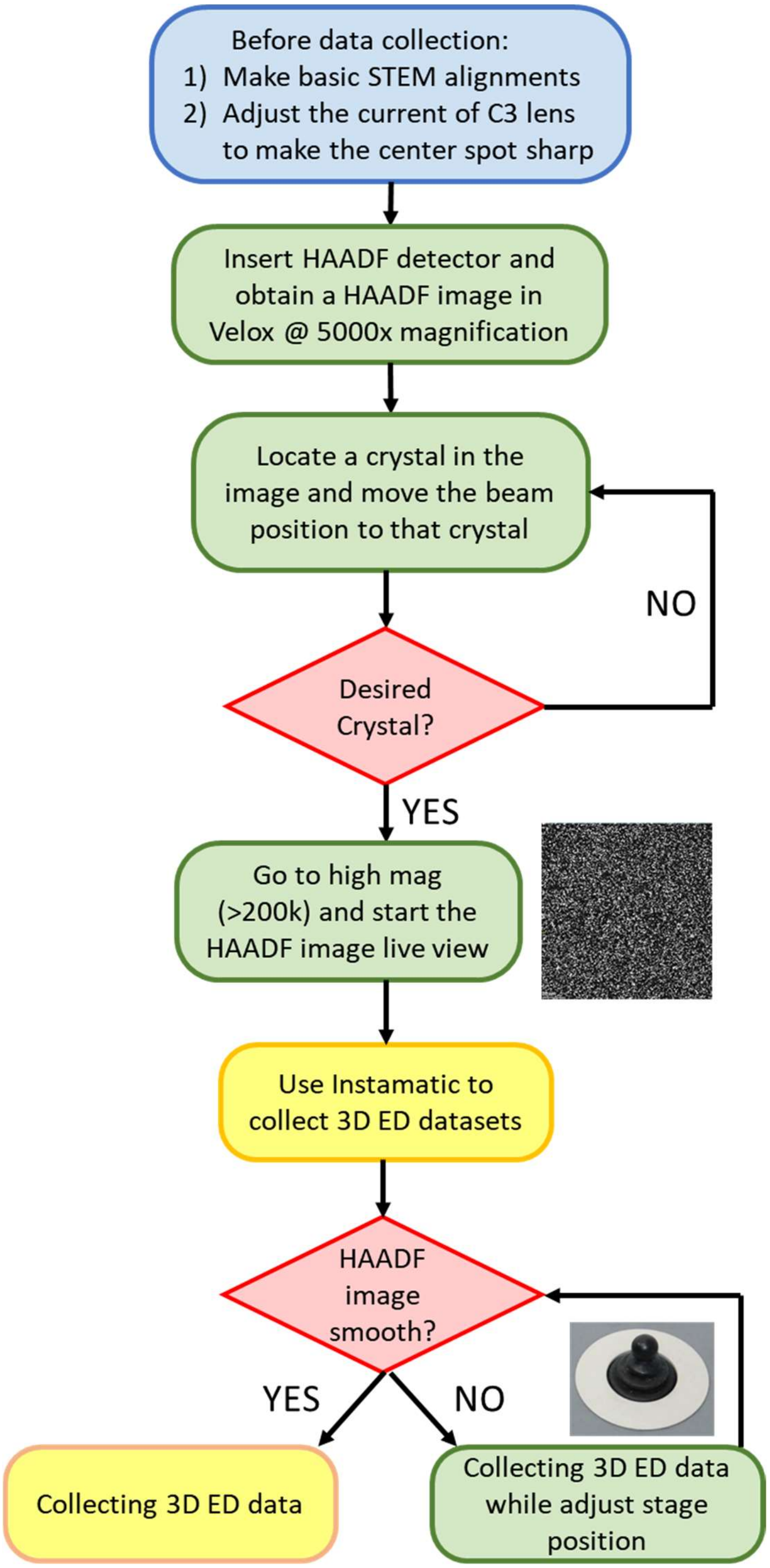

Figure 2. The workflow for STEM HAADF crystal tracking while collecting 3DED datasets using *Instamatic*. The blue, green and red boxes show steps that requires human intervention. The yellow boxes show steps of Instamatic protocol running automatically. The STEM image at low magnification shows the positions where might find a desired crystal. The STEM image at high magnification shows a snapshot when collecting 3DED datasets.

Table 1. Data processing details using XDS and crystallographic details for the refinement for eight datasets of $NH_4H_2PO_4$ with and without energy filtration.

| **Dataset no.** | **1** | **2** | **3** | **4** | **5** | **6** |
|---|---|---|---|---|---|---|
| Energy-filtered | Yes | Yes | Yes | No | No | No |
| Rotation Range (°) | 144.3 | 140.5 | 137.0 | 130.9 | 140.9 | 143.1 |
| Resolution (Å) | 0.85 | 0.85 | 0.85 | 0.85 | 0.85 | 0.85 |
| *a* (Å) | 7.59 | 7.59 | 7.58 | 7.56 | 7.57 | 7.37 |
| *b* (Å) | 7.62 | 7.62 | 7.61 | 7.55 | 7.59 | 7.43 |
| *c* (Å) | 7.65 | 7.69 | 7.62 | 7.59 | 7.68 | 7.71 |
| α (°) | 90.44 | 90.72 | 91.14 | 90.69 | 90.82 | 90.68 |
| β (°) | 90.94 | 91.34 | 90.88 | 90.19 | 91.21 | 91.89 |
| γ (°) | 91.36 | 91.44 | 91.20 | 91.43 | 91.30 | 89.33 |
| No. of reflections (Fo > 4sig(Fo)) | 174 | 173 | 166 | 171 | 166 | 155 |
| No. of reflections (all unique) | 189 | 188 | 183 | 179 | 184 | 183 |
| $R_1$ (Fo > 4sig(Fo)) | 8.5% | 10.5% | 11.9% | 11.0% | 11.6% | 14.5% |
| $R_1$ (all reflections) | 8.9% | 12.6% | 12.7% | 11.1% | 14.2% | 18.9% |
| Goof | 1.409 | 1.158 | 1.193 | 1.240 | 1.352 | 1.442 |
| No. of Parameters | 21 | 21 | 21 | 17 | 21 | 21 |
| No. of restraints | 0 | 0 | 0 | 0 | 0 | 0 |

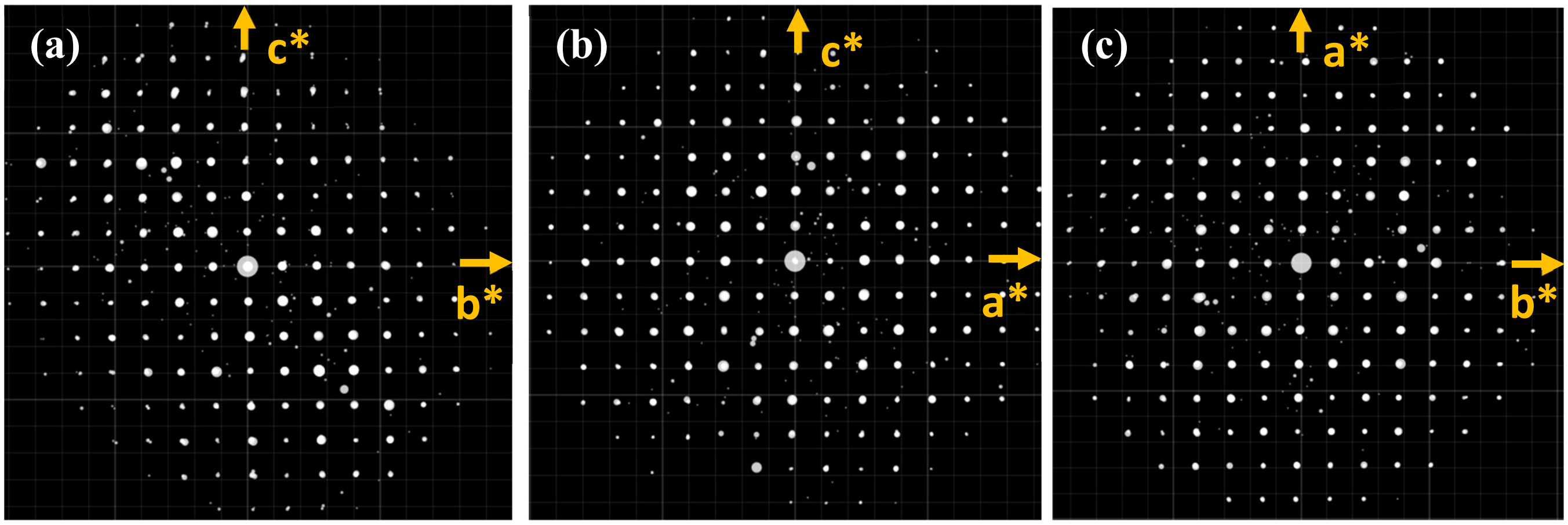

Figure 3. Typical 3D reciprocal lattices of $NH_4H_2PO_4$ collected on GIF reconstructed and visualized by REDp software (Wan et al., 2013).

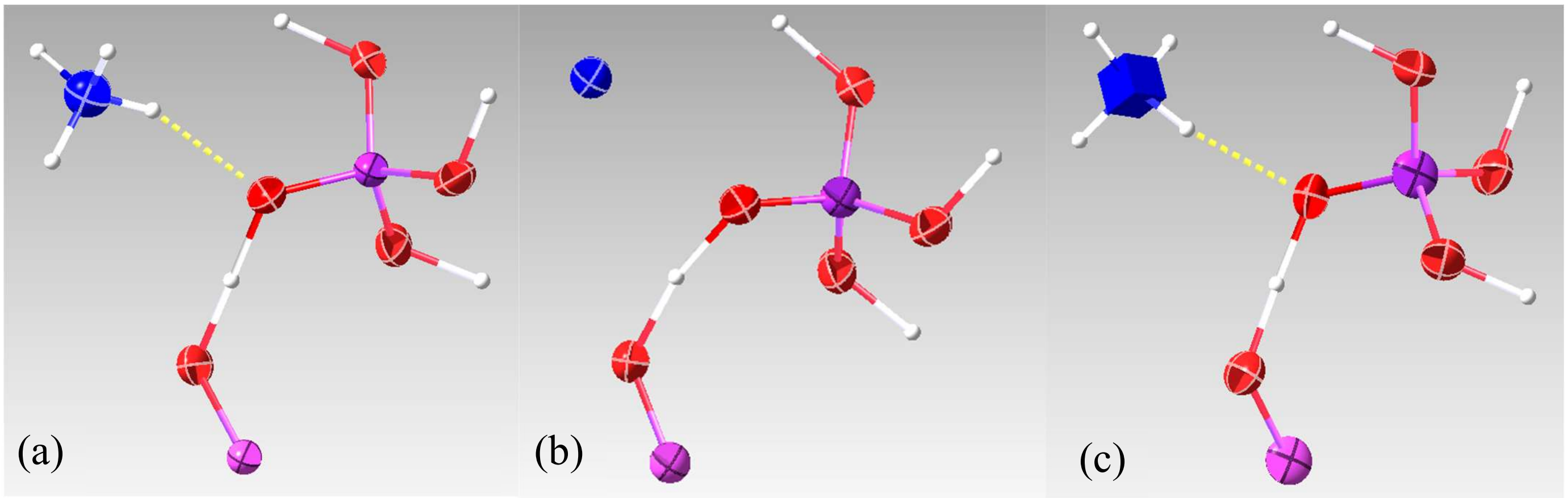

Figure 4. (a) A typical $NH_4H_2PO_4$ crystal structure representation from a filtered dataset. All hydrogen atoms were found and all the ADPs are reasonable after anisotropic refinement (b, c) Two $NH_4H_2PO_4$ crystal structure representations from unfiltered datasets. The refinement was not able to find all the hydrogen atoms around nitrogen atom or unable to obtain reason thermal parameters for the nitrogen atom. The dotted line represents the ionic bond between hydrogen atom and oxygen atom. White for hydrogen; blue for nitrogen; purple for phosphate and red for oxygen.

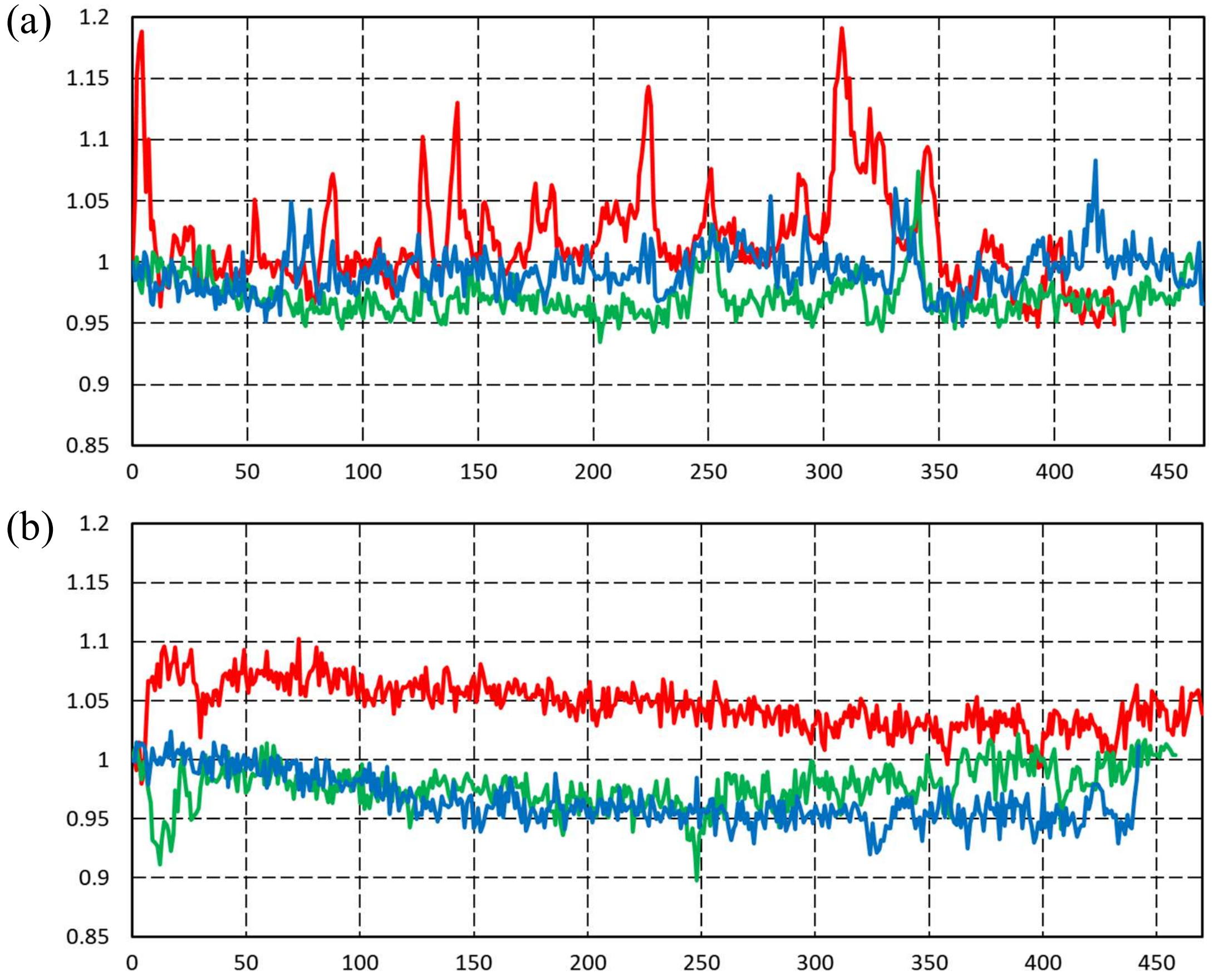

Figure 5. Normalized scaling factors from diffraction patterns collected for (a) three filtered datasets and (b) three unfiltered datasets of $NH_4H_2PO_4$ as calculated by XDS (SCALE in file INIT.Lp) can be used to judge the tracking of the crystals. If the crystal moves (partially) out of the SA aperture, the image scale is affected.

Table 2. Data processing details using XDS and crystallographic details for the refinement for eight datasets of ZSM-5 with and without energy filtration.

| **Dataset no.** | **1** | **2** |
|---|---|---|
| Energy-filtered | No | Yes |
| Rotation Range (°) | 121.0 | 124.2 |
| Resolution (Å) | 0.90 | 0.90 |
| Completeness (%) | 82.4 | 84.5 |
| I/SIGMA | 3.53 | 4.34 |
| CC(1/2) (%) | 98.8 | 99.6 |
| Observed Reflections | 12439 | 12546 |
| R-meas (%) | 22.5 | 15.9 |
| No. of reflections (Fo > 4sig(Fo)) | 1702 | 1775 |
| No. of reflections (all unique) | 3279 | 3354 |
| $R_1$ (Fo > 4sig(Fo)) | 26.4% | 24.3% |
| $R_1$ (all reflections) | 30.6% | 28.7% |
| Goof | 1.619 | 1.740 |
| No. of Parameters | 332 | 332 |
| No. of restraints | 0 | 0 |

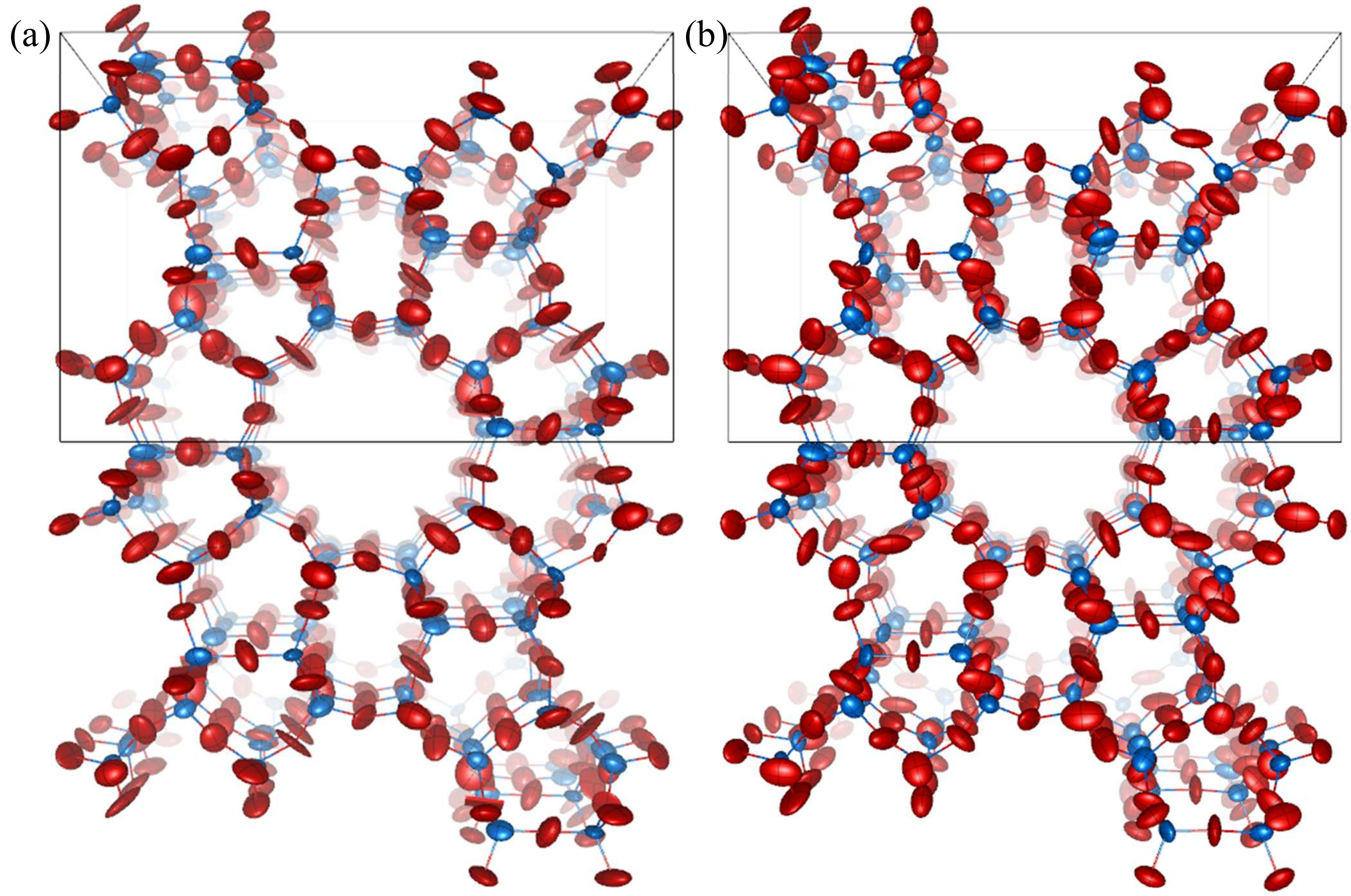

Figure 6. Refined structures of ZSM-5 from (a) unfiltered dataset (b) filtered dataset, showing atomic displacement parameters for the Si and O atoms at the 60% probability level along *b* axis. Red for oxygen atoms and blue for silicon atoms. The structure from unfiltered dataset contained a lot of unreasonable ADPs. Some of them became negative ADPs while the ADPs of structure from filtered dataset were reasonable and closer to isotropic state.

# Supplementary Information

## Improving data quality for 3D electron diffraction (3DED) by Gatan Image Filter and a new crystal tracking method

**Taimin Yang, Hongyi Xu, Xiaodong Zou**

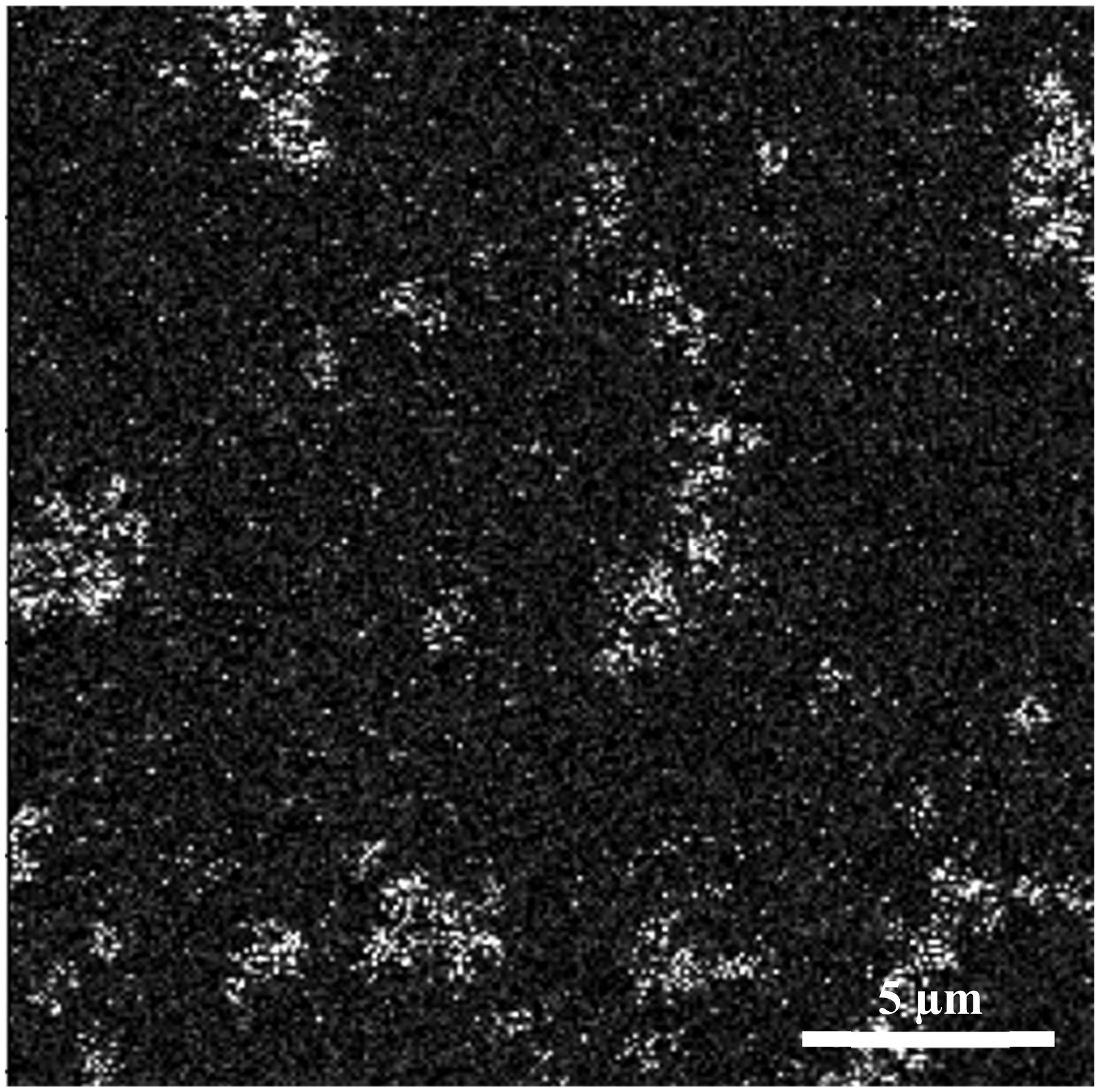

Figure S1. Typical HAADF image of ZSM-5 sample at 5000x magnification with a parallel beam of 300nm in diameter. The bright parts indicate possible ZSM-5 crystal targets.

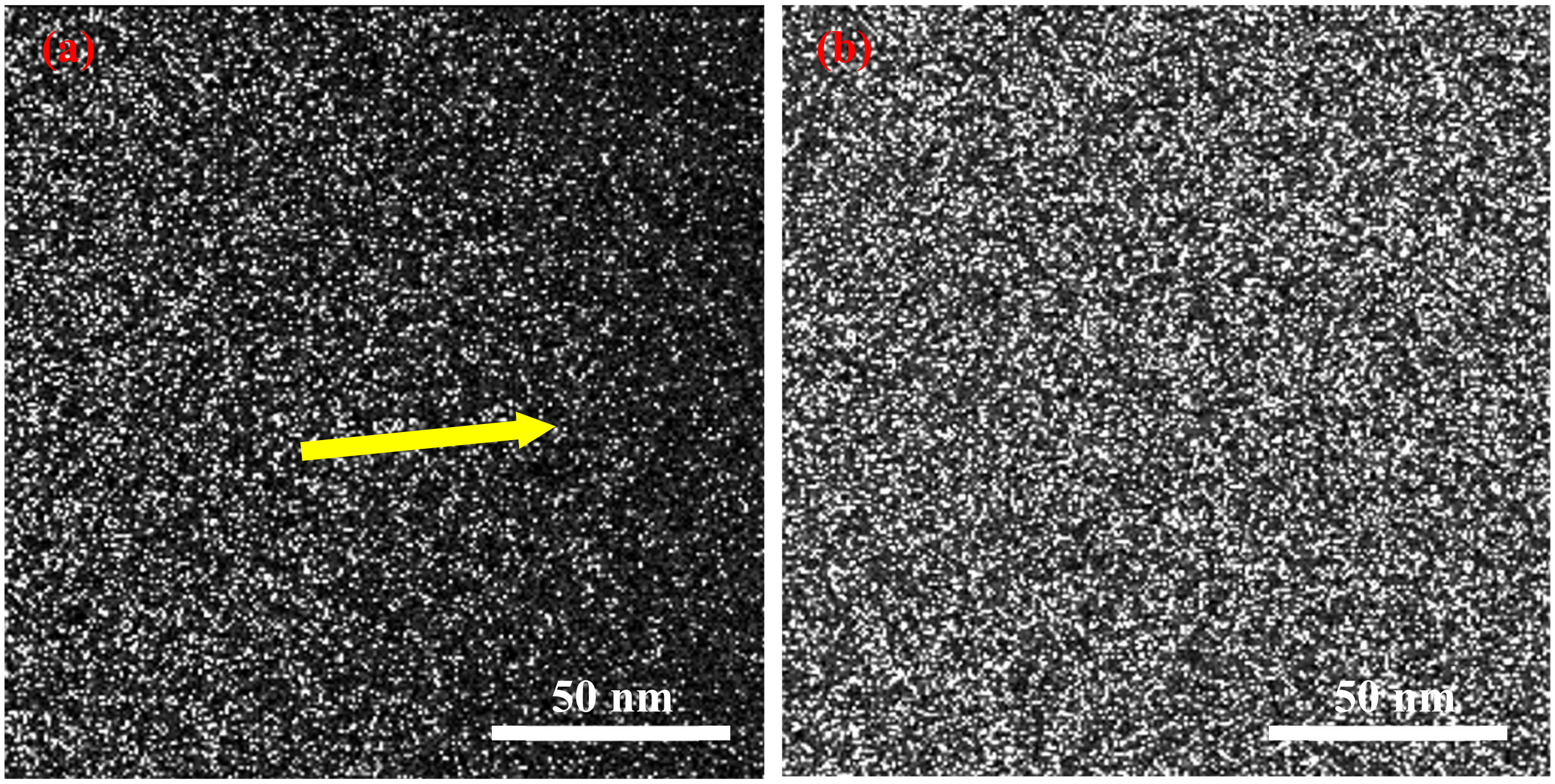

Figure S2. Typical HAADF images of ZSM-5 sample at high magnification during 3DED data collection. (a) An example when scanning area was located at the edge of the target crystal. The position of the darker area in the image indicates the stage adjustment direction, as shown by the blue arrow. (b) After stage adjustment, the scanning area moved back to the center part of the crystal.

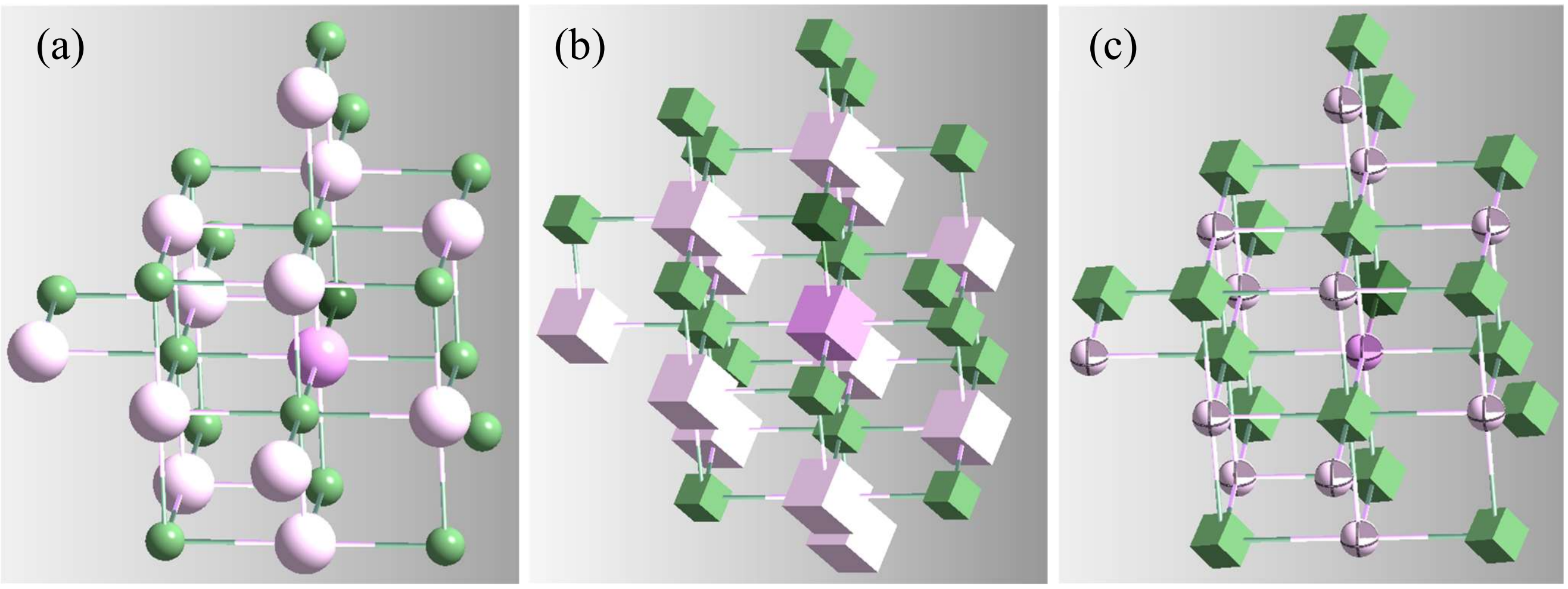

Figure S3. Typical structure of NaCl refined (a) with EXTI keyword (b) without EXTI keyword (c) with SWAT keyword.

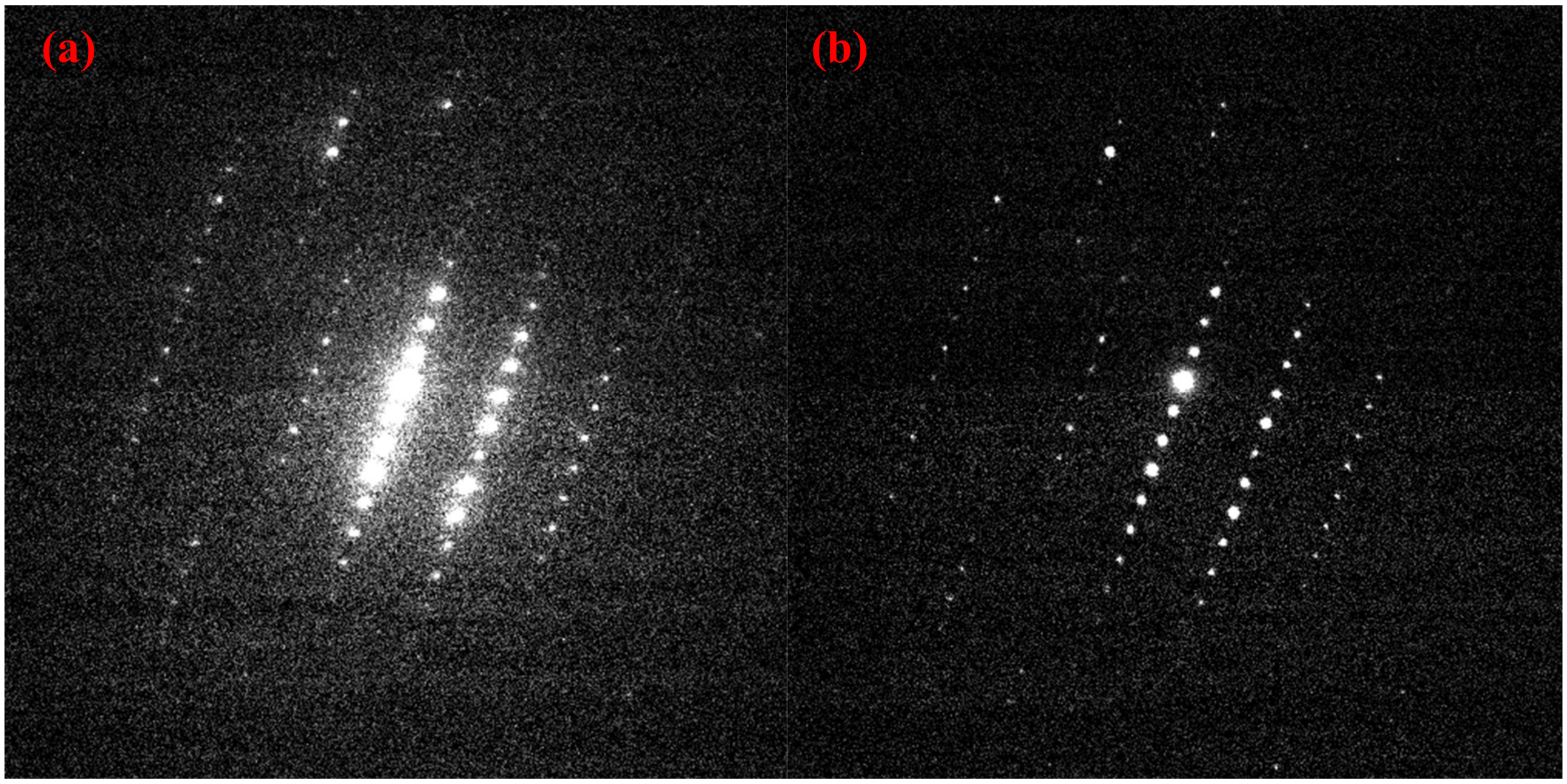

Figure S4. Typical electron diffraction patterns of ZSM-5 crystal (a) without energy filtration and (b) with energy filtration. The profiles of the reflections in the filtered electron diffraction pattern were much sharper.

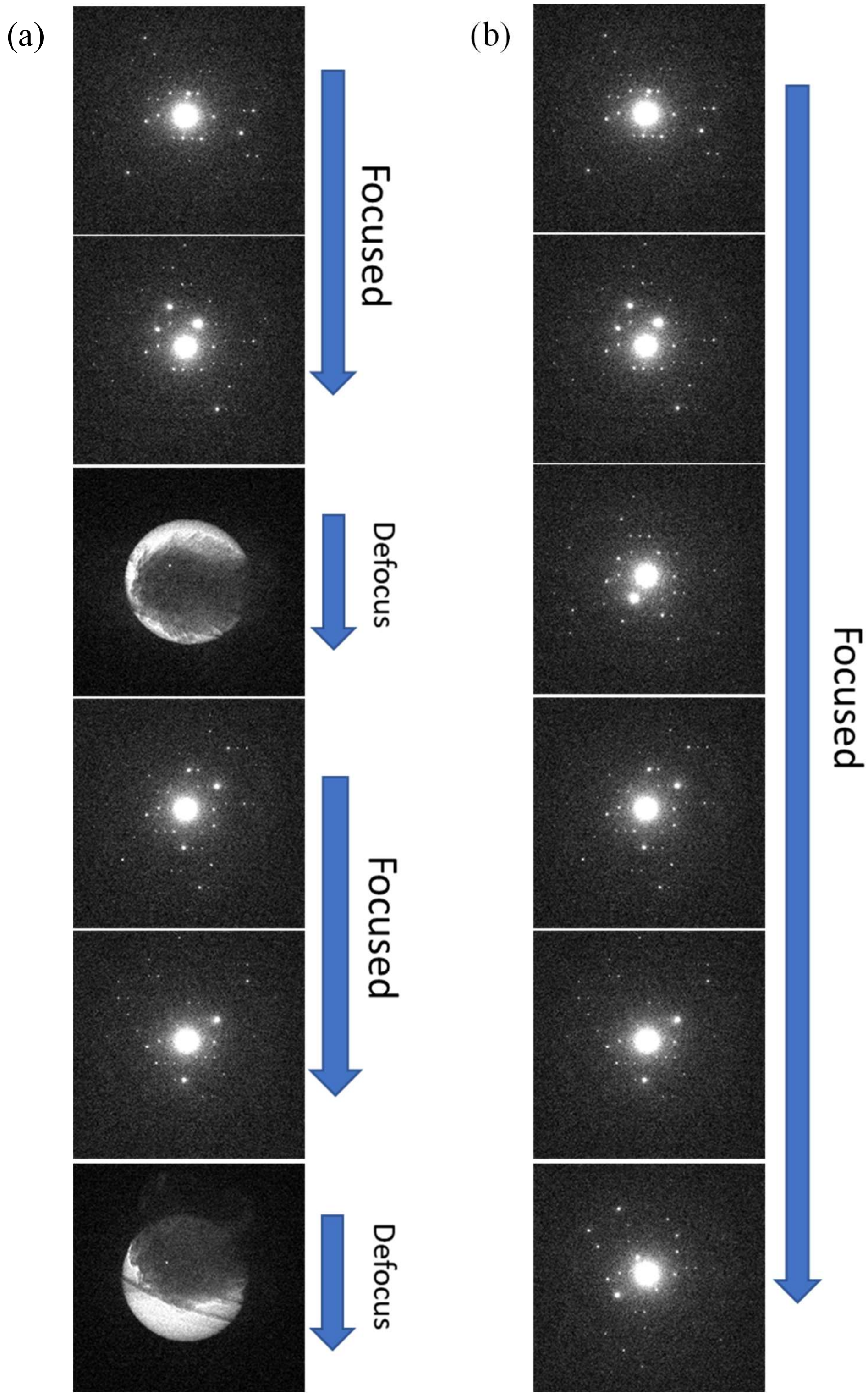

Figure S5. Comparison between different crystal tracking method (a) defocus diffraction pattern tracking (b) live STEM-HAADF imaging tracking

Table S1. Data processing details using XDS and crystallographic details for the refinement for eight datasets of NaCl collected in STEM mode with EXTI keyword.

| **Dataset no.** | **1** | **2** | **3** | **4** | **5** | **6** | **7** | **8** |
|---|---|---|---|---|---|---|---|---|
| Energy-filtered | Yes | Yes | Yes | Yes | No | No | No | No |
| Rotation Range (°) | 137.2 | 150.4 | 150.6 | 133.4 | 136.1 | 144.3 | 135.2 | 137.9 |
| Resolution (Å) | 0.8 | 0.8 | 0.8 | 0.8 | 0.8 | 0.8 | 0.8 | 0.8 |
| *a* (Å) | 5.52 | 5.54 | 5.53 | 5.54 | 5.53 | 5.68 | 5.65 | 5.67 |
| *b* (Å) | 5.61 | 5.54 | 5.56 | 5.67 | 5.62 | 5.62 | 5.65 | 5.68 |
| *c* (Å) | 5.79 | 5.81 | 5.85 | 5.71 | 5.80 | 5.73 | 5.72 | 5.65 |
| α (°) | 92.18 | 91.86 | 90.17 | 92.86 | 91.71 | 90.18 | 89.97 | 91.35 |
| β (°) | 89.51 | 90.61 | 91.45 | 91.43 | 89.73 | 90.40 | 90.12 | 90.95 |
| γ (°) | 90.82 | 89.68 | 89.90 | 88.99 | 91.61 | 91.61 | 91.81 | 90.93 |
| No. refl (Fo>4sig(Fo)) | 21 | 19 | 21 | 21 | 21 | 18 | 19 | 16 |
| No. all unique refl | 21 | 20 | 21 | 21 | 21 | 19 | 20 | 18 |
| $R_1$ (Fo > 4sig(Fo)) | 7.9% | 7.5% | 10.1% | 8.3% | 11.7% | 13.8% | 15.6% | 14.7% |
| $R_1$ (all reflections) | 7.9% | 7.9% | 10.1% | 8.3% | 11.7% | 15.2% | 15.7% | 15.9% |
| Goof | 1.34 | 1.461 | 1.464 | 1.217 | 1.34 | 1.383 | 1.481 | 1.038 |
| No. of Parameters | 4 | 4 | 4 | 4 | 4 | 4 | 4 | 4 |

Table S2. The refinement results for eight datasets of NaCl collected in STEM mode without EXTI keyword.

| **Dataset no.** | **1** | **2** | **3** | **4** | **5** | **6** | **7** | **8** |
|---|---|---|---|---|---|---|---|---|
| No. refl (Fo>4sig(Fo)) | 21 | 19 | 21 | 21 | 21 | 18 | 19 | 16 |
| No. all unique refl | 21 | 20 | 21 | 21 | 21 | 19 | 20 | 18 |
| $R_1$ (Fo > 4sig(Fo)) | 32.2% | 21.0% | 23.6% | 20.8% | 33.5% | 19.8% | 23.5% | 32.6% |
| $R_1$ (all reflections) | 32.2% | 21.2% | 23.6% | 20.8% | 33.5% | 19.8% | 24.2% | 30.8% |
| Goof | 1.026 | 1.432 | 1.276 | 1.242 | 1.076 | 1.323 | 1.322 | 2.151 |
| No. of Parameters | 3 | 3 | 3 | 3 | 3 | 3 | 3 | 3 |

Table S3. The refinement results for eight datasets of NaCl collected in STEM mode with SWAT keyword. The refinement for dataset 5 was unstable so the results were omitted for that dataset.

| **Dataset no.** | **1** | **2** | **3** | **4** | **5** | **6** | **7** | **8** |
|---|---|---|---|---|---|---|---|---|
| No. refl (Fo>4sig(Fo)) | 21 | 19 | 21 | 21 | 21 | 18 | 19 | 16 |
| No. all unique refl | 21 | 20 | 21 | 21 | 21 | 19 | 20 | 18 |
| $R_1$ (Fo > 4sig(Fo)) | 23.5% | 19.0% | 20.1% | 21.8% | / | 19.5% | 20.7% | 30.4% |
| $R_1$ (all reflections) | 23.5% | 19.3% | 20.1% | 21.8% | / | 22.0% | 21.3% | 29.1% |
| Goof | 1.159 | 1.517 | 2.470 | 2.275 | / | 2.410 | 2.652 | 2.151 |
| No. of Parameters | 5 | 5 | 5 | 5 | 5 | 5 | 5 | 5 |

Table S4. Deviations of atomic positions between the reference ZSM-5 structure (van Koningsveld *et al.*, 1987) and those determined from filtered and unfiltered 3DED data collected in STEM modes. Fractional atomic coordinates for the reference ZSM-5 structure determined by SCXRD (as-made ZSM-5, space group *Pnma*, $a$ = 20.022(4) Å, $b$ = 19.899(4) Å, $c$ = 13.383(3) Å, see the International Zeolite Association (IZA) Database).

| Atom | Atom displacement unfiltered, Å | Atom displacement filtered, Å |
|---|---|---|
| Si1 | 0.025 | 0.028 |
| Si2 | 0.036 | 0.049 |
| Si3 | 0.061 | 0.033 |
| Si4 | 0.055 | 0.019 |
| Si5 | 0.051 | 0.040 |
| Si6 | 0.048 | 0.018 |
| Si7 | 0.057 | 0.066 |
| Si8 | 0.044 | 0.023 |
| Si9 | 0.043 | 0.025 |
| Si10 | 0.061 | 0.058 |
| Si11 | 0.025 | 0.046 |
| Si12 | 0.066 | 0.023 |
| O1 | 0.088 | 0.092 |
| O2 | 0.070 | 0.031 |
| O3 | 0.050 | 0.070 |
| O4 | 0.119 | 0.104 |
| O5 | 0.030 | 0.068 |
| O6 | 0.128 | 0.088 |
| O7 | 0.102 | 0.045 |
| O8 | 0.127 | 0.146 |
| O9 | 0.053 | 0.031 |
| O10 | 0.135 | 0.124 |
| O11 | 0.051 | 0.115 |
| O12 | 0.140 | 0.147 |
| O13 | 0.093 | 0.045 |
| O14 | 0.040 | 0.069 |
| O15 | 0.116 | 0.103 |
| O16 | 0.079 | 0.110 |
| O17 | 0.086 | 0.055 |
| O18 | 0.102 | 0.046 |
| O19 | 0.031 | 0.073 |
| O20 | 0.115 | 0.022 |
| O21 | 0.087 | 0.076 |
| O22 | 0.089 | 0.152 |
| O23 | 0.092 | 0.064 |
| O24 | 0.088 | 0.040 |
| O25 | 0.028 | 0.046 |
| O26 | 0.058 | 0.065 |
| **<Si> average** | 0.048(13) | 0.035(15) |
| **<O> average** | 0.085(33) | 0.078(37) |

Table S5. Refined Si-O bond distances for ZSM-5 structure from filtered and unfiltered and the reference structure. The number marked with red color means severe deviation from reference bond length.

| Atomic 1 | Atomic 2 | Bond length unfiltered, Å | Bond length filtered, Å | Bond length ref, Å |
|---|---|---|---|---|
| SI1 | O1 | 1.6091 | 1.5863 | 1.5830 |
| SI1 | O15 | **1.4953** | 1.6079 | 1.5914 |
| SI1 | O16 | 1.5976 | 1.5778 | 1.5800 |
| SI1 | O21 | 1.6438 | 1.6188 | 1.5977 |
| SI2 | O1 | 1.6031 | 1.5701 | 1.5867 |
| SI2 | O2 | 1.6261 | 1.5654 | 1.6011 |
| SI2 | O6 | 1.5593 | 1.6620 | 1.5816 |
| SI2 | O13 | **1.4943** | 1.6422 | 1.5676 |
| SI3 | O2 | 1.6064 | 1.6198 | 1.5867 |
| SI3 | O3 | 1.6068 | 1.5706 | 1.5708 |
| SI3 | O19 | 1.5732 | 1.5021 | 1.5711 |
| SI3 | O20 | 1.5144 | 1.6056 | 1.5914 |
| SI4 | O3 | 1.5355 | 1.5813 | 1.5748 |
| SI4 | O4 | 1.6345 | 1.6661 | 1.5861 |
| SI4 | O16 | 1.5596 | 1.6203 | 1.5825 |
| SI4 | O17 | 1.5809 | 1.6421 | 1.5889 |
| SI5 | O4 | 1.5530 | 1.5016 | 1.5829 |
| SI5 | O5 | 1.5774 | 1.5436 | 1.5891 |
| SI5 | O14 | 1.5709 | 1.6330 | 1.5825 |
| SI5 | O21 | 1.5547 | 1.5113 | 1.5980 |
| SI6 | O5 | 1.6274 | 1.6686 | 1.5942 |
| SI6 | O6 | 1.6007 | 1.5377 | 1.5879 |
| SI6 | O18 | 1.5713 | 1.6239 | 1.5935 |
| SI6 | O19 | 1.5528 | 1.6236 | 1.5855 |
| SI7 | O7 | 1.5626 | 1.5221 | 1.5804 |
| SI7 | O17 | 1.5816 | 1.5108 | 1.5860 |
| SI7 | O23 | 1.5947 | 1.6347 | 1.5849 |
| SI7 | O22 | 1.5683 | 1.5132 | 1.5905 |
| SI8 | O7 | 1.5931 | 1.5921 | 1.5856 |
| SI8 | O8 | 1.5791 | 1.5685 | 1.5882 |
| SI8 | O12 | **1.6819** | 1.5776 | 1.5827 |
| SI8 | O13 | 1.6344 | 1.5172 | 1.5765 |
| SI9 | O8 | 1.6201 | 1.6187 | 1.5783 |
| SI9 | O9 | 1.5844 | 1.6015 | 1.5909 |
| SI9 | O25 | 1.5591 | 1.5917 | 1.5983 |
| SI9 | O18 | 1.6628 | 1.5921 | 1.5971 |
| SI10 | O9 | 1.6285 | 1.6096 | 1.5897 |
| SI10 | O10 | 1.6081 | 1.5967 | 1.5730 |
| SI10 | O26 | 1.6259 | 1.6017 | 1.6049 |
| SI10 | O15 | 1.6313 | 1.5471 | 1.5884 |
| SI11 | O10 | 1.5329 | 1.5724 | 1.5809 |
| SI11 | O11 | 1.5983 | 1.6361 | 1.5910 |
| SI11 | O14 | 1.5788 | 1.5588 | 1.5681 |
| SI11 | O22 | 1.5560 | 1.6798 | 1.5937 |
| SI12 | O11 | 1.6006 | 1.5055 | 1.5857 |
| SI12 | O12 | **1.4469** | 1.5580 | 1.5742 |
| SI12 | O20 | 1.6148 | 1.6054 | 1.6055 |
| SI12 | O24 | 1.6086 | 1.6240 | 1.5952 |
| **Average Length** | | **1.5863(476)** | **1.5868(442)** | **1.5866(88)** |

Table S6. Data processing details using XDS for two datasets of ZSM-5 collected in TEM mode with and without energy filtration

| | Unfiltered | Energy-filtered |
|---|---|---|
| Observed Reflections | 9199 | 9920 |
| I/SIGMA | 7.74 | 9.25 |
| CC(1/2) | 99.7 | 99.8 |
| Rotation Range (°) | 64.6 | 69.8 |
| Resolution (Å) | 0.9 | 0.9 |
| *a* (Å) | 19.76 | 19.75 |
| *b* (Å) | 19.73 | 19.67 |
| *c* (Å) | 13.66 | 13.67 |
| α (°) | 91.57 | 91.00 |
| β (°) | 90.85 | 90.23 |
| γ (°) | 90.32 | 91.68 |